\documentclass{article}
\usepackage{spconf,amsmath,graphicx,hyperref}
\usepackage{booktabs} 
\usepackage{color}
\usepackage{colortbl}
\usepackage{amssymb}
\usepackage{enumitem}
\usepackage{xcolor}

\title{Recovering Performance in Speech Emotion Recognition from Discrete Tokens via Multi-Layer Fusion and Paralinguistic Feature Integration}

\name{Esther Sun, Abinay Reddy Naini, Carlos Busso}
\address{Language Technologies Institute, Carnegie Mellon University, USA\\
esthers@andrew.cmu.edu, anaini@andrew.cmu.edu, busso@cmu.edu}
\begin{document}
\ninept
\maketitle
\begin{abstract}
Discrete speech tokens offer significant advantages for storage and language model integration, but their application in \emph{speech emotion recognition} (SER) is limited by paralinguistic information loss during quantization. This paper presents a comprehensive investigation of discrete tokens for SER. Using a fine-tuned WavLM-Large model, we systematically quantify performance degradation across different layer configurations and k-means quantization granularities. To recover the information loss, we propose two key strategies: (1) attention-based multi-layer fusion to recapture complementary information from different layers, and (2) integration of openSMILE features to explicitly reintroduce paralinguistic cues. We also compare mainstream neural codec tokenizers (SpeechTokenizer, DAC, EnCodec) and analyze their behaviors when fused with acoustic features. Our findings demonstrate that through multi-layer fusion and acoustic feature integration, discrete tokens can close the performance gap with continuous representations in SER tasks.
\end{abstract}
\section{Introduction}
\label{sec:intro}

Self-supervised learning (SSL) models, such as Wav2Vec2 \cite{Baevski_2020}, WavLM \cite{Chen_2022}, and HuBERT \cite{Hsu_2021}, have recently achieved remarkable performance in numerous audio and speech processing tasks by learning intricate, continuous representations from raw waveforms \cite{Wagner_2023,Goncalves_2024,Naini_2025_2}. Meanwhile, significant research effort has emerged to convert these continuous features into discrete audio tokens. This paradigm shift is driven by the potential of discrete units for highly efficient data transmission and storage, their native compatibility with token-based architectures such as modern \emph{large language models} (LLMs) \cite{Yang_2023, Yang_2025, Tang_2023}, and their potential for cross-modal alignment \cite{Liu_2024, Cheng_2024}. However, quantization inevitably introduces information loss, making a key challenge in the area of speech tokenization to preserve as much relevant information as possible while converting continuous waveforms into compact discrete token sequences.

Mainstream speech tokenization approaches can be broadly categorized into two strategies: 
(1) clustering the hidden embeddings from specific Transformer-Encoder layers of an SSL model using algorithms such as K-Means \cite{Polyak_2021, Wells_2022}, and (2) relying on neural audio codecs, which originally designed for audio reconstruction. Discrete tokens derived from SSL models have proven highly effective for linguistic tasks \cite{Chang_2024} such as \emph{automatic speech recognition} (ASR), translation, and understanding, suggesting their significant potential for \emph{speech emotion recognition} (SER) tasks that rely heavily on phonetic information. However, the majority of existing studies focus exclusively on clustering the final layer of SSL models \cite{Li_2025, Mote_2025}, or only select a limited number of layers and codebook size for simple analysis \cite{Mousavi_2024_2}, neglecting an in-depth systematic investigation into the contributions of intermediate layers or the impact of varying cluster quantities on SER performance. Moreover, current evaluations of discrete tokens for SER are often relegated to general downstream task benchmarks \cite{Mousavi_2024, Mousavi_2025}. There is a notable lack of in-depth analysis regarding which specific aspects of information, particularly paralinguistic and prosodic cues, are lost during the discretization process when applying discrete tokens to SER tasks. More critically, it is not clear how to effectively recover this lost information to maintain competitive performance.

This paper addresses these gaps by conducting a comprehensive evaluation of discrete tokens for an 8-class SER task. Unlike most prior work, we utilize a fine-tuned WavLM model as the source for our discrete representations. We systematically compare multiple-layer configurations and k-means cluster sizes, quantifying performance sensitivities that arise from discretization. Recognizing that discrete features may not adequately preserve the paralinguistic and prosodic information \cite{Nguyen_2023, Ren_2024}, which is essential for SER tasks, we employ attention-based cross-layer fusion to combine information across WavLM layers architecturally, and augment discrete tokens with seven families of discretized OpenSMILE features at the feature level, thereby complementing discrete tokens with  paralinguistic/acoustic features. Additionally, we compare mainstream neural codecs such as SpeechTokenizer \cite{Zhang_2023_ST}, DAC\cite{Kumar_2023_DAC}, and EnCodec\cite{Defossez_2022}, as alternative tokenizers. We study their fusion with openSMILE features in the same 8-class SER regime. To summarize, our contributions in this work are: 
\begin{itemize}[leftmargin=1.2em]
\vspace{-1em}
\setlength{\itemindent}{1em}
\setlength{\itemsep}{0cm}%
\setlength{\parskip}{0cm}%
\item \textbf{Layer-wise, granularity-controlled analysis.} We provide a systematic analysis of discrete tokens for SER by clustering each layer of a fine-tuned WavLM-large model with K-Means, evaluating the performance across various layer configurations and cluster quantities.
\item \textbf{Feature- and architecture-level compensation.} We mitigate discretization losses by applying attention-based cross-layer fusion and integrating seven openSMILE feature families to recover the lost information.
\item \textbf{Neural-codec comparison and fusion.} We compared SpeechTokenizer, DAC, and EnCodec as tokenizers for a 8-class SER task and evaluate their fusion with openSMILE features.
\end{itemize}



\section{Method}
\label{sec:pagestyle}

Emotion information is compromised in the speech discretization process. We explored two key ideas to recover emotional information in discrete speech units. After presenting our proposed base architecture (Sec. \ref{ssec:base-architecture}), we present our strategies based on attention-based multi-layer fusion (Sec. \ref{ssec:attention}) and representation enhancement with discrete paralinguistic features (Sec. \ref{ssec:paralinguistic}).

\subsection{Proposed SSL Model Architecture}
\label{ssec:base-architecture}

\begin{figure}[t]
\begin{minipage}[b]{1.0\linewidth}
  \centering
  \centerline{\includegraphics[width=0.98\linewidth]{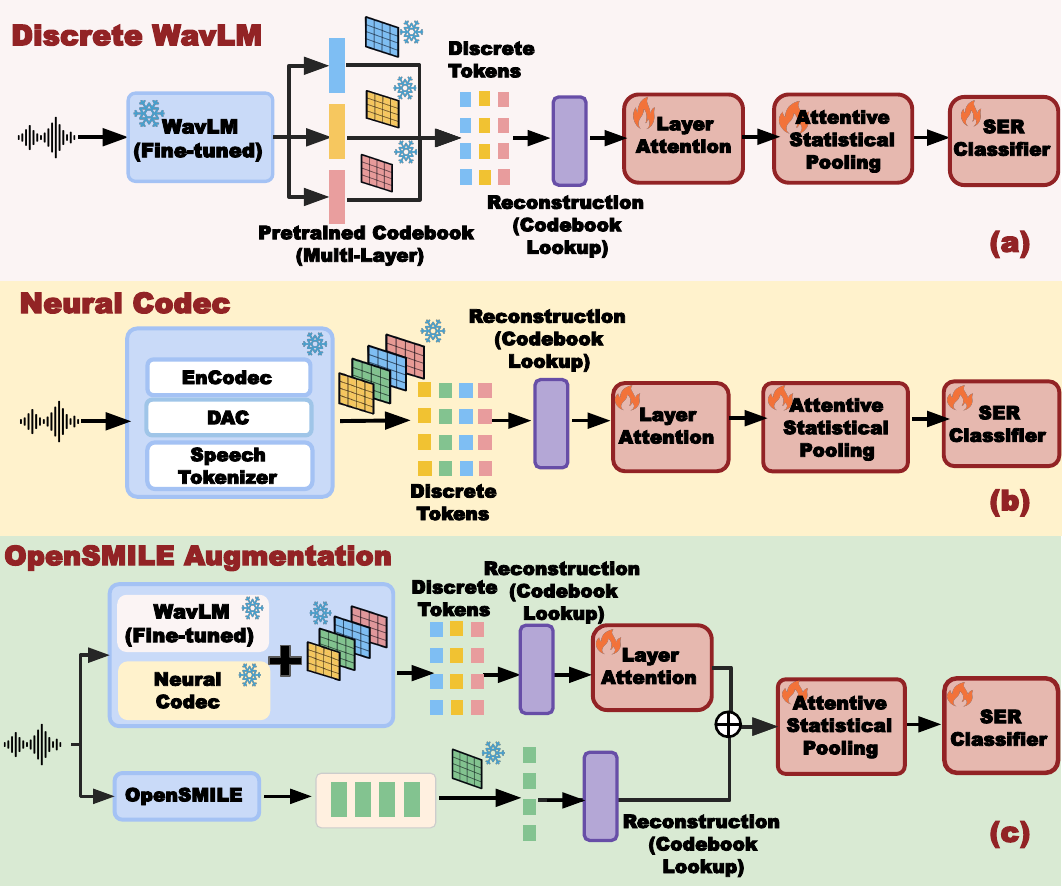}}
\end{minipage}
\caption{Proposed discrete SER framework. (a) Multi-layer discrete units from fine-tuned WavLM with layer-specific codebooks. (b) Neural codec tokenization via EnCodec, DAC, and SpeechTokenizer. (c) Hierarchical fusion augmenting discrete speech representations with quantized OpenSMILE paralinguistic features. All models share the same downstream architecture (layer attention, pooling, classifier).}
\label{fig:res}
\end{figure}

As shown in the top of Figure \ref{fig:res}(a), our proposed SER architecture learns to combine multi-layer discrete features derived from a frozen, pre-trained and fine-tuned SSL model. The pipeline (i) extracts hidden states from selected layers, (ii) quantizes them via nearest-centroid assignment to per-layer K-Means codebooks, (iii) reconstructs discrete features from code indices, (iv) fuses multi-layer information by a temperature-controlled attention mechanism, and (v) performs attentive statistics pooling followed by a lightweight classification head.

Using a frozen \textsc{WavLM-Large} model fine-tuned on the MSP-Podcast dataset, we extract hidden-state representations $H^{(\ell)} \in \mathbb{R}^{T \times D}$ from a set of specified Transformer layers $\mathcal{L} \subseteq \{0, \dots, 23\}$. We then discretize these features via vector quantization. For each layer $\ell \in \mathcal{L}$, we create a layer-specific codebook $C^{(\ell)} \in \mathbb{R}^{K \times D}$ by applying k-means clustering to its features with $K \in \{256, 512, 1{,}000, 2{,}000, 4{,}000\}$. Each frame vector is subsequently mapped to its nearest centroid index $z_t^{(\ell)}$, and its discrete representation is reconstructed via centroid lookup: $\tilde{H}_t^{(\ell)} = C^{(\ell)}_{z_t^{(\ell)}}$. These codebooks are generated once for all 24 layers and then fixed for all subsequent experiments to produce the discrete speech units.

We also explore neural audio codecs \textbf{(SpeechTokenizer, DAC, EnCodec)} as alternative tokenizers for SER (Figure \ref{fig:res}(b)). These models generate multi-layer discrete representations directly from raw waveforms via learned encoders and \emph{residual vector quantization} (RVQ). For the SER task, we extract multi-layer discrete representations from their encoding pathway, which are then processed using the exact same downstream architecture as the SSL model (comprising layer attention, pooling, and classification modules).


\subsection{Attention-Based Multi-Layer Fusion}
\label{ssec:attention}

Studies have shown that different layers provide varying levels of emotional information \cite{Pepino_2021}. We hypothesize that relying on discretized units from different layers can be beneficial in bridging the gap with continuous representation. To integrate multi-layer discrete representations, we employ attention-based fusion instead of simple averaging. For $N$ selected layers, a learnable module computes dynamic weights $\{\alpha_1, \dots, \alpha_N\}$ through a masked average pooling and a temperature-scaled softmax mechanism with learnable temperature $\tau$. We apply layer normalization before the attention mechanism to prevent scale variations from affecting the attention weights. The fused representation $H_{\text{fused}} = \sum_{l \in \mathcal{L}} \alpha_l \cdot \hat{H}_l$ is processed by an attentive statistics pooling layer and a \emph{multilayer perceptron} (MLP) classifier. Only these downstream modules are fine-tuned with a weighted cross-entropy loss while freezing the SSL extractor.


\subsection{Enhancement via Discrete Paralinguistic Features}
\label{ssec:paralinguistic}
To compensate for information loss due to discretization, we design a method that fuses discrete tokens with discrete speech units derived from traditional paralinguistic acoustic features (Figure \ref{fig:res}(c)). In particular, we utilize the low-level features from the \emph{Computational Paralinguistics Challenge} (ComParE) and the extended Geneva Minimalistic Acoustic Parameter Set (eGeMAPS) \cite{Eyben_2016} from OpenSMILE, which comprise 74-dimensional features spanning seven acoustic categories (Table~\ref{tab:opensmile_features}). This feature set has been extensively used in different paralinguistic tasks. To maintain consistency with the tokenized representation space, we quantize OpenSMILE features into discrete tokens using the K-Means algorithm, with codebook size $K$ determined by the elbow method. Our hierarchical fusion strategy first performs attention-based fusion on multi-layer discrete representations: for SSL models (e.g., WavLM), these are discrete units from different transformer layers; for neural audio codecs (e.g., EnCodec, DAC, SpeechTokenizer), these are tokens from different RVQ codebook layers. The fused representation yields $H_{\text{fused}}$. The discretized OpenSMILE features are temporally aligned to match the target sequence length through adaptive resampling (linear interpolation for upsampling or uniform sampling for downsampling), forming $H_{\text{OpenSMILE}}$. A learnable modality normalizer then applies LayerNorm and scaling parameters $(\gamma_{\text{fused}}, \gamma_{\text{OpenSMILE}})$ to each modality to balance their contributions before concatenation. This unified approach effectively combines the semantic richness of tokenized speech representations with explicit paralinguistic cues, improving emotion classification performance.

\section{Experimental Setup}
\label{sec:exp}

All experiments are conducted on the MSP-Podcast corpus \cite{Busso_2025, Lotfian_2019_3} (v1.12) for an 8-class emotion recognition task (anger, sadness, happiness, surprise, fear, disgust, contempt, and neutral speech). The MSP-Podcast corpus is a large-scale naturalistic emotional speech database. Version 1.12 contains 207,136 utterances from speech recordings, providing diverse speakers and realistic emotional expressions in spontaneous scenarios. We use the Macro F1 Score as the evaluation metric to account for class imbalance. Each experiment is conducted three times, and we report the average results to mitigate the impact of randomness. 

To analyze the effect of codebook size and layer configuration on SER, we experiment with five cluster counts, $K \in \{256, 512, 1{,}000, 2{,}000, 4{,}000\}$. Each $K$-value is evaluated across six representative layer configurations: \textbf{All Layers} (L0-L23), \textbf{All but Last Layer} (L0-L22), \textbf{Last Layer Only} (L23), \textbf{Sparse Layers} (L1,3,7,12,18,23), the \textbf{Last 8 Layers} (L16-L23), and a \textbf{10-Layer} set (L0,1,2,4,6,9,12,16,20,23).

\begin{table}[t]
\centering
\caption{OpenSMILE Feature Categories (each category is discretized with the K-Means algorithm, using a codebook of size $k$).}
\label{tab:opensmile_features}
\resizebox{\columnwidth}{!}{%
\begin{tabular}{lccl}
\hline
\textbf{Category} & \textbf{Dim} & \textbf{k} & \textbf{Description} \\
\hline
Prosody & 6 & 32 & Pitch, energy, voicing \\
Spectral & 14 & 64 & Band energies, moments \\
MFCC & 14 & 64 & Cepstral coefficients \\
Voice Quality & 5 & 32 & Jitter, shimmer, HNR \\
Formants & 6 & 32 & F1-F3 frequencies \\
Auditory Bands & 26 & 128 & 26-band energy \\
Additional & 3 & 16 & ZCR, sharpness \\
\hline
\textbf{Total} & \textbf{74} & \textbf{416} & \textbf{7 categories} \\
\hline
\end{tabular}%
}
\end{table}


We also evaluate three mainstream neural audio codecs as alternative tokenizers with their encoders frozen. Specifically, we use \textbf{SpeechTokenizer} (official 16kHz model) with representations from its 2th, 4th, and 8th codebook layers; \textbf{DAC} (24kHz model) with representations from its 2nd, 4th, and 9th codebook layers; and 3 \textbf{EnCodec} variants with various bandwidths and quantization layers: \texttt{bw\_3.0kbps} (4 layers), \texttt{bw\_6.0kbps} (8 layers), and \texttt{bw\_24.0kbps} (32 layers).

We augment discrete OpenSMILE features to discrete tokens from both SSL models and neural codecs. For the discrete WavLM tokens ($K$=1{,}000), augmentation is tested on four representative layer configurations: \textbf{Last 8 Layers, Last Layer Only, L0-L23, and Sparse Layers.} For the neural codecs, it is applied to all configurations previously detailed.

\begin{figure}[t]
\begin{minipage}[b]{1.0\linewidth}
  \centering
  \centerline{\includegraphics[width=8.5cm]{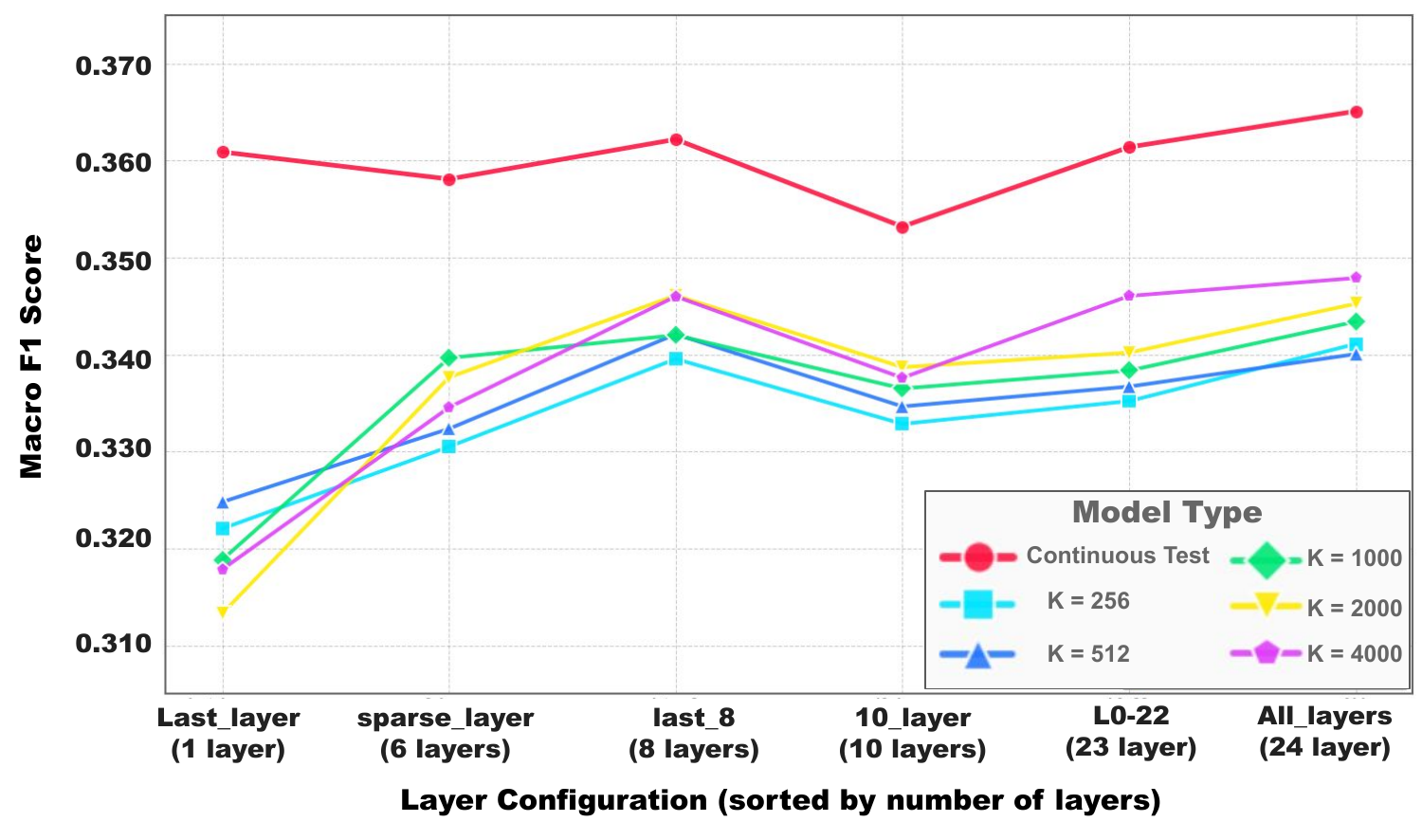}}
\end{minipage}
\caption{Performance (Macro F1 Score) across different WavLM layer configurations and codebook sizes (K). The continuous feature-based model (red line) is included as a reference baseline, using the corresponding layers.} 
\label{fig:line_plot}
\end{figure}

\begin{table*}[t]
\centering
\caption{Comprehensive SER performance (Macro F1 Score) comparison. The table evaluates discrete models (WavLM, neural codecs) and the impact of augmenting them with various paralinguistic feature sets. Discrete WavLM models consistently outperform neural codecs, and their performance is further boosted by selective feature augmentation.}
\label{tab:feature_comparison}
\resizebox{\textwidth}{!}{%
\begin{tabular}{lccccccccc}
\hline
\textbf{Model} & \textbf{None} & \textbf{Prosody} & \textbf{Voice} & \textbf{MFCC} & \textbf{Spectral} & \textbf{Formants} & \textbf{Auditory} & \textbf{Additional} & \textbf{All Features} \\
 & \textbf{(baseline)} &  & \textbf{Quality} &  &  &  & \textbf{Bands} &  & \textbf{(74)} \\
\hline
Speech Tokenizer (ST-2) & 0.1486 & 0.1299 & 0.1313 & 0.1416 & \textbf{0.1567} & 0.1384 & 0.1344 & 0.1366 & 0.1510 \\
Speech Tokenizer (ST-4) & 0.1619 & 0.1214 & 0.1616 & 0.1447 & 0.1598 & 0.1432 & 0.1295 & 0.1485 & 0.1413 \\
Speech Tokenizer (ST-8) & 0.1758 & 0.1437 & 0.1681 & 0.1403 & 0.1533 & 0.1550 & 0.1397 & 0.1667 & 0.1675 \\
\hline
DAC (0,1) & 0.1011 & 0.1172 & 0.1129 & 0.1243 & 0.1147 & \textbf{0.1421} & 0.1149 & 0.1083 & 0.1432 \\
DAC (0,1,2,3) & 0.1187 & 0.1476 & 0.1144 & \textbf{0.1520} & 0.1141 & 0.1476 & 0.1412 & 0.1413 & 0.1423 \\
DAC [0-8] & 0.1159 & \textbf{0.1438} & 0.1166 & 0.1253 & 0.1066 & 0.1089 & 0.1244 & 0.1071 & 0.1376 \\
\hline
Encodec (bw\_3.0kbps\_4\_layers) & 0.1575 & 0.1669 & 0.1702 & 0.1690 & 0.1720 & 0.1677 & 0.1743 & 0.1557 & \textbf{0.1755} \\
Encodec (bw\_6.0kbps\_8\_layers) & 0.1599 & 0.1855 & 0.1531 & \textbf{0.1857} & 0.1624 & 0.1638 & 0.1662 & 0.1549 & 0.1728 \\
Encodec (bw\_24.0kbps\_32\_layers) & 0.1550 & 0.1893 & 0.1323 & 0.1909 & 0.1541 & 0.1827 & 0.1363 & 0.1723 & \textbf{0.2005} \\
\hline
Discrete WavLM (L1,3,7,12,18,23) & 0.3371 & 0.3482 & 0.3436 & 0.3401 & \textbf{0.3501} & 0.3493 & 0.3447 & 0.3411 & 0.3498 \\
Discrete WavLM (L16-23) & 0.3420 & 0.3463 & 0.3439 & 0.3478 & 0.3497 & 0.3440 & 0.3440 & 0.3431 & \textbf{0.3505} \\
Discrete WavLM (L0-23) & 0.3441 & 0.3452 & 0.3467 & 0.3479 & 0.3493 & \textbf{0.3534} & 0.3483 & 0.3442 & 0.3461 \\
Discrete WavLM (L23) & 0.3120 & 0.3123 & 0.3126 & 0.3145 & \textbf{0.3183} & 0.3131 & 0.3101 & 0.3114 & 0.3126 \\
\hline
\end{tabular}
}
\vspace{-0.2cm}
\end{table*}

\vspace{-0.3cm}
\section{Findings}
\label{sec:majhead}
\vspace{-0.2cm}
\subsection{Impact of Discretization}
To analyze the impact of speech token discretization from the fine-tuned WavLM on SER performance, we tested six representative layer configurations and five codebook sizes (see Section \ref{sec:exp}) for both continuous features and discrete tokens using the corresponding layers. Experimental results reveal that the discretization process leads to significant performance degradation: discrete tokens achieve Macro F1 scores ranging from 0.3133 to 0.3479, representing a 6--14\% performance drop compared to continuous features. Notably, continuous features demonstrate stable performance across all layer configurations, consistently achieving Macro F1 scores around 0.36, even when using only the last layer (Macro F1 = 0.362). This stability contrasts sharply with discrete token fluctuations, highlighting information loss in paralinguistic cues during quantization.

We also evaluated neural audio codecs (SpeechTokenizer, EnCodec, DAC) on SER tasks. As shown in Table~\ref{tab:feature_comparison}, these codecs exhibit severe performance degradation: SpeechTokenizer achieves the highest codec performance (Macro F1 = 0.1758 with ST-8), followed by EnCodec (Macro F1 = 0.1599) and DAC (Macro F1 = 0.1187). Similar findings are reported by \cite{Mousavi_2025}, where discrete WavLM, even unfinetuned, consistently outperforms SpeechTokenizer, EnCodec, and DAC for an SER task on the IEMOCAP corpus.  Compared to discrete fine-tuned WavLM (minimum Macro F1 = 0.3133), neural audio codecs suffer over 50\% performance degradation, indicating their representations are optimized for speech reconstruction rather than emotion-relevant features. These results emphasize the importance of selecting appropriate discretization methods for SER tasks.

\vspace{-0.2cm}
\subsection{Attention-based Multi-Layer Fusion}
Despite performance degradation from discretization, our multi-layer fusion strategy demonstrates significant compensatory effects. Single-layer discrete features (Layer 23) plateau at Macro F1 = 0.3248, while combining all 24 layers with $K = 4{,}000$ achieves Macro F1 = 0.3479, recovering approximately 62.1\% of the discretization-induced performance gap. This result indicates earlier layers retain complementary emotional information lost in single-layer discretization.

\begin{figure}[t]
\begin{minipage}[b]{1.0\linewidth}
  \centering
  \centerline{\includegraphics[width=8.5cm]{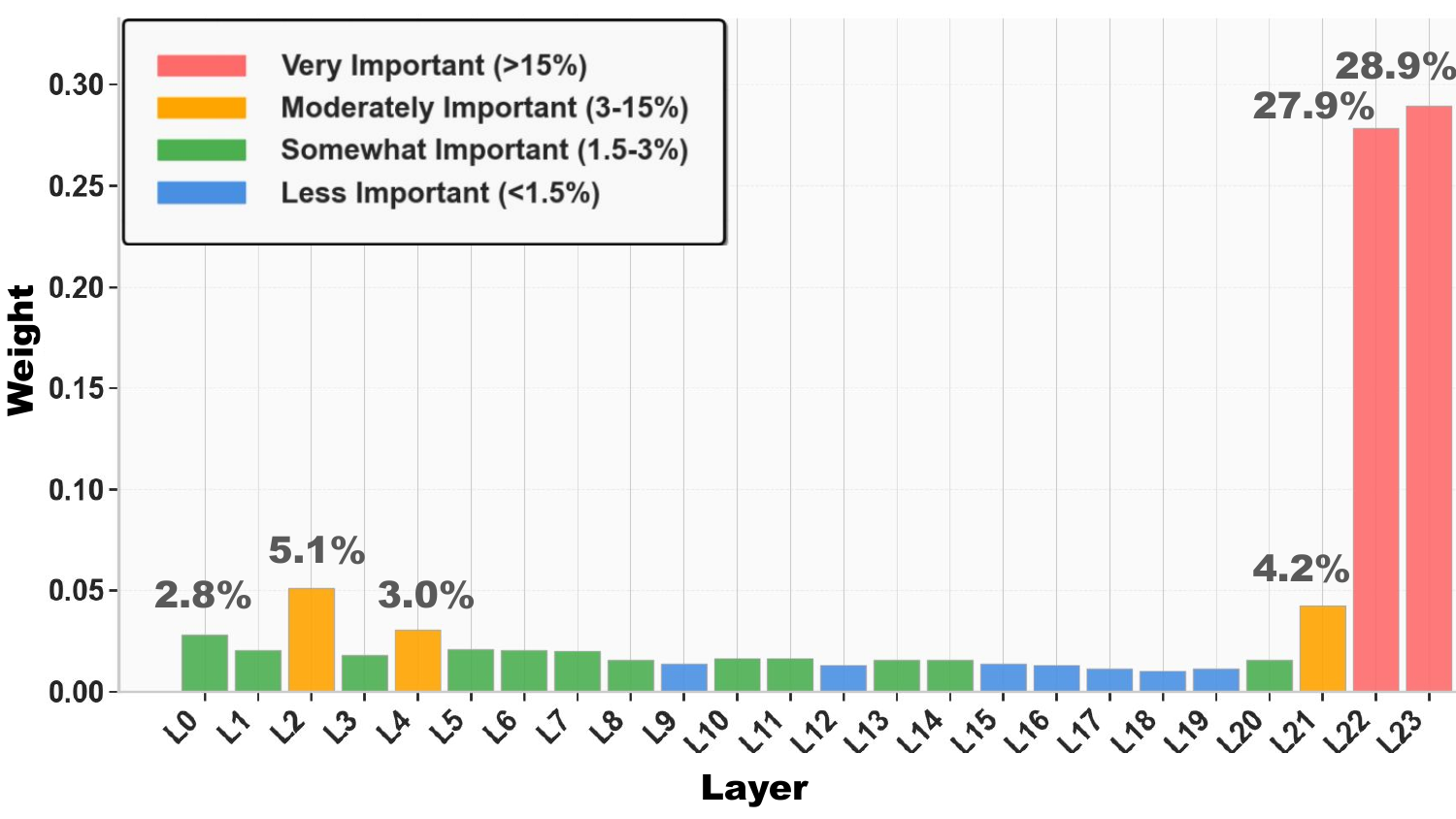}}
\end{minipage}
\caption{Analysis of attention weights across all 24 WavLM layers, $K = 1{,}000$, Macro F1 = 0.3441. The distribution is notably bimodal: the final two layers (L22, L23) are critically important,  while select early layers also make contributions.}
\label{fig:layer_weights}
\end{figure}

Figure~\ref{fig:layer_weights} reveals a bimodal attention weight distribution after layer normalization: the final two layers dominate (L22: 27.9\%, L23: 28.9\%, totaling approximately 57\%), while earlier layers receive non-trivial attention. This result suggests the model primarily leverages high-level semantic features while selectively incorporating low-level acoustic information for emotion recognition. Layer selection strategy proves more important than layer quantity. The last-8-layer configuration (Layers 16--23) achieves Macro F1 = 0.346, outperforming the 10-layer configuration (Macro F1 = 0.3376) despite fewer layers. Layer configuration also correlates with optimal codebook size: sparse configurations favor smaller codebooks ($K = 256$--$512$), while dense ones benefit from larger codebooks. However, gains beyond $K = 1{,}000$ are marginal (only 0.59\% improvement when increasing to $K = 4{,}000$).

Notably, multi-layer fusion provides minimal benefits for neural audio codecs. Table~\ref{tab:feature_comparison} shows SpeechTokenizer improves only marginally from 2 to 8 layers (ST-2: 0.1486 to ST-8: 0.1758), while DAC and EnCodec degrade with more layers. DAC drops performance from 0.1187 (4-layer) to 0.1159 (8-layer), and EnCodec from 0.1599 to 0.1550. This contrasts sharply with discrete WavLM's 10.3\% improvement through multi-layer fusion, confirming that neural audio codec representations primarily serve speech reconstruction rather than  preservation of paralinguistic information.

\begin{figure}[t]
\begin{minipage}[b]{1.0\linewidth}
  \centering
  \centerline{\includegraphics[width=8.5cm]{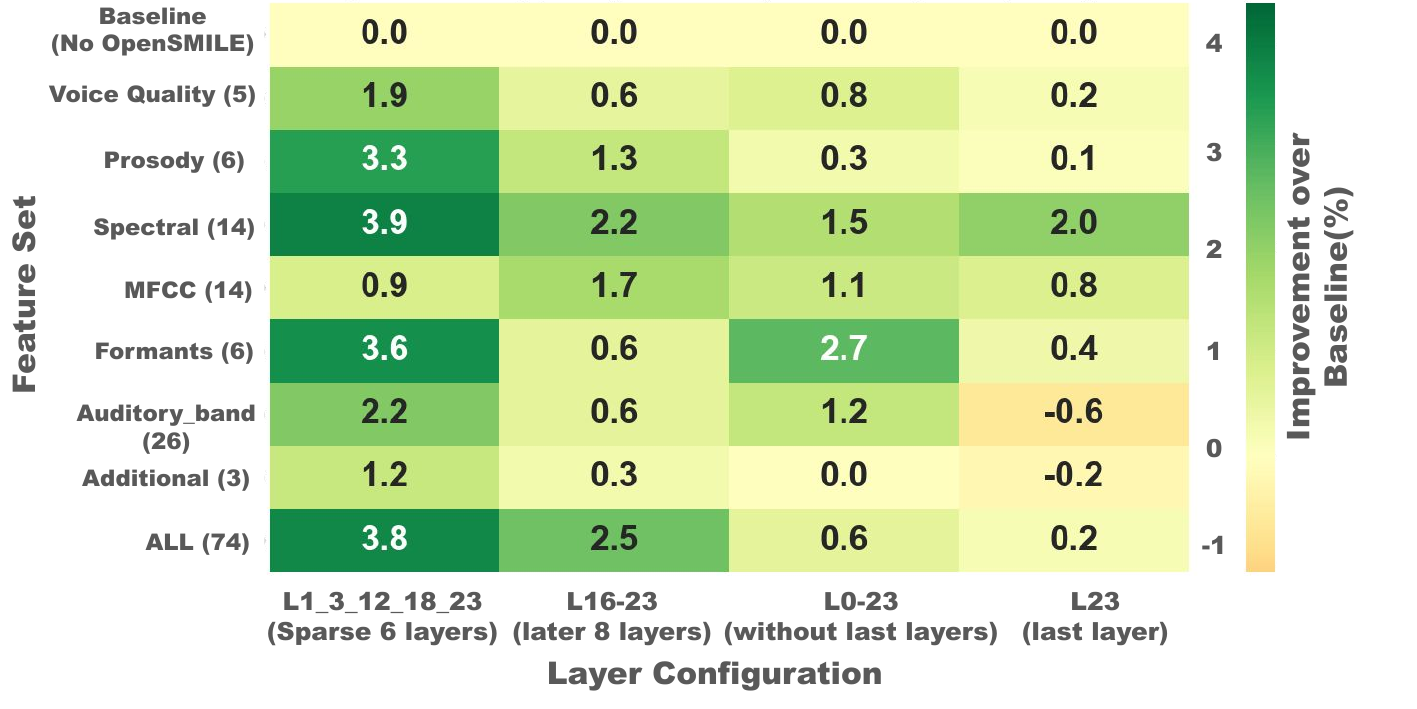}}
\end{minipage}
\vspace{-0.7cm}
\caption{Performance improvement (\%) from augmenting discrete WavLM models ($K = 1{,}000$) with paralinguistic features. The results show a clear inverse relationship: sparser layer configurations (left) benefit the most from explicit paralinguistic cues, while denser configurations that already contain richer information (right) show diminishing returns.}

\label{fig:performance_heatmap}
\end{figure}

\vspace{-0.2cm}
\subsection{Adding Discrete Tokens and Paralinguistic Features}
Table~\ref{tab:feature_comparison} and Figure~\ref{fig:performance_heatmap} show the performance when combining discrete speech tokens with the discretized paralinguistic features. The discrete WavLM tokens show varying degrees of improvement with paralinguistic feature augmentation, revealing a clear inverse relationship between layer density and augmentation benefit. The sparse layer selection (L1,3,7,12,18,23) consistently demonstrates the strongest gains across most feature types (averaging 2--4\% improvement), followed by L16--23 with moderate improvements (1--2.5\%). The dense L0--23 configuration shows minimal gains (0--2.7\%) despite having all layers' information. The single L23 layer, capturing task-specific information from fine-tuned WavLM, exhibits the weakest response (mostly under 1\%). Among individual features, spectral features boost performance for the sparse layers by 3.9\% and even benefit L23 by 2.0\%. Formants achieve the highest single gain for L0--23 (2.7\%) and sparse layers (3.6\%). Prosody features also notably enhance sparse configurations (3.3\%) but have minimal impact on denser ones. This pattern suggests that sparse representations, missing intermediate acoustic-prosodic information due to layer sampling, can effectively leverage explicit paralinguistic features to fill these gaps. The proposed augmentation proves valuable for recovering discretization losses. While a SER model trained with continuous WavLM achieves Macro F1 = 0.3624, our best augmented configuration (L0--23 with Formants) reaches Macro F1 = 0.3534, recovering approximately 75\% of the performance gap and demonstrating the effectiveness of paralinguistic features in bridging the discrete-continuous divide.

For neural codecs, acoustic augmentation shows inconsistent trends, with most configurations showing modest improvements (Table~\ref{tab:feature_comparison}). SpeechTokenizer generally degrades with augmentation except for ST-2 with Spectral features (5.5\% gain). DAC shows the most dramatic selective improvements: DAC (0,1) gains 40.6\% with Formants and DAC (0,1,2,3) improves 28.0\% with MFCC. The trends for EnCodec are more consistent, where all the configurations benefit from the augmentation, particularly bw\_24.0kbps, achieving 29.4\% with all features. Notably, Formants, MFCC, and the complete feature sets provide the most substantial enhancements across codecs. This pattern suggests that codec representations, optimized for reconstruction rather than semantic content, have specific acoustic gaps that complementary features can effectively fill. However, the benefit is highly codec-dependent, unlike the consistent gains seen in SSL-based tokens.

\vspace{-0.3cm}
\section{Conclusion}
\vspace{-0.2cm}
This paper systematically studied discrete tokens for SER through multi-layer SSL analysis, openSMILE fusion, and neural codec comparison. Key findings: (1) Multi-layer fusion of discrete WavLM tokens significantly outperforms single-layer discrete tokens, recovering 75\% of SER quantization loss; (2) OpenSMILE features effectively restore paralinguistic information (Macro F1 = 0.3534); (3) Neural codecs underperform and resist acoustic augmentation, unlike SSL tokens. The results demonstrate that strategic multi-layer fusion with acoustic compensation enables high-performance discrete SER.

\bibliographystyle{IEEEbib}
\bibliography{reference}

@article{Busso_2025,
    	author = {C. Busso and R. Lotfian and  K. Sridhar and A.N. Salman and W.-C. Lin and L. Goncalves and S. Parthasarathy and A. {Reddy Naini} and S.-G. Leem and L. Martinez-Lucas and H.-C. Chou and P. Mote},
    	title = {The MSP-Podcast Corpus},
    	journal = {ArXiv e-prints (arXiv:2509.09791)},
    	archivePrefix = {arXiv},
    	eprint = {2509.09791},
    	primaryClass = {eess.AS},
    	volume = {},
    	number = {},
    	year = {2025},
    	pages = {1-20},
    	month = {September},
    	doi={10.48550/arXiv.2509.09791},
}

@InProceedings{Naini_2025_2, 
	author={Reddy Naini, A. and others}, 
	title={The {Interspeech} 2025 Challenge on Speech Emotion Recognition in Naturalistic Conditions},
	booktitle={Interspeech 2025}, 
	volume={},
	year={2025}, 
	month={August}, 
	address = {Rotterdam, The Netherlands},
	pages={4668-4672}, 
	doi={10.21437/Interspeech.2025-1972},
}

@InProceedings{Mote_2025, 
	author={P. Mote and A. {Reddy Naini} and D. Robinson and E. Richerson and C. Busso}, 
	title={Analysis of Phonetic Level Similarities Across Languages in Emotional Speech},
	booktitle={Interspeech 2025}, 
	volume={},
	year={2025}, 
	month={August}, 
	address =  {Rotterdam, The Netherlands},
	pages={}, 
	doi={10.21437/Interspeech.2025-2112},
}

@InProceedings{Goncalves_2024, 
	author={Goncalves, L. and others}, 
	title={Odyssey 2024 - Speech Emotion Recognition Challenge: Dataset, Baseline Framework, and Results},
	booktitle={The Speaker and Language Recognition Workshop (Odyssey 2024)}, 
	volume={},
	year={2024}, 
	month={June}, 
	address = {Quebec, Canada},
	pages={247-254}, 
	doi={10.21437/odyssey.2024-35},
}

@article{Wagner_2023, 
    author = {Wagner, J. and others}, 
    title = {Dawn of the Transformer Era in Speech Emotion Recognition: Closing the Valence Gap}, 
    journal = {IEEE Transactions on Pattern Analysis and Machine Intelligence}, 
    volume = {45}, 
    number = {9}, 
    year = {2023}, 
    pages = {10745-10759}, 
    month = {September}, 
    doi={10.1109/TPAMI.2023.3263585}
}

@article{Chen_2022,
	author = {S. Chen and others},
	title = {{WavLM}: Large-Scale Self-Supervised Pre-Training for Full Stack Speech Processing},
	journal = {IEEE Journal of Selected Topics in Signal Processing},
	volume = {16},
	number = {6},
	year = {2022},
	pages = {1505-1518},
	month = {October},
	doi={10.1109/JSTSP.2022.3188113},
}

@article{Hsu_2021,
	author = {W.-N. Hsu and B. Bolte, Y.-H. H. Tsai and K. Lakhotia and R. Salakhutdinov and A. Mohamed},
	title = {{HuBERT}: Self-Supervised Speech Representation Learning by Masked Prediction of Hidden Units},
	journal = {IEEE/ACM Transactions on Audio, Speech, and Language Processing},
	volume = {29},
	number = {},
	year = {2021},
	pages = {3451-3460},
	month = {},
	doi={10.1109/TASLP.2021.3122291},
}

@InProceedings{Pepino_2021, 
	author={L. Pepino and P. Riera and L. Ferrer}, 
	title={Emotion Recognition from Speech Using {Wav2vec} 2.0 Embeddings},
	booktitle={Interspeech 2021}, 
	volume={},
	year={2021}, 
	month={August-September}, 
	address =  {Brno, Czech Republic},
	pages={3400-3404}, 
	doi={10.21437/Interspeech.2021-703},
}

@InProceedings{Baevski_2020, 
	author={A. Baevski and Y. Zhou and A. Mohamed and M. Auli}, 
	title={Wav2vec 2.0: A Framework for Self-Supervised Learning of Speech Representations},
	booktitle={Advances in Neural Information Processing Systems (NeurIPS 2020)}, 
	volume={33},
	year={2020}, 
	month={December}, 
	address =  {Virtual},
	pages={12449-12460}, 
	doi={},
}

@article{Lotfian_2019_3,
	author = {R. Lotfian and C. Busso},
	title = {Building Naturalistic Emotionally Balanced Speech Corpus by Retrieving Emotional Speech From Existing Podcast Recordings},
	journal = {IEEE Transactions on Affective Computing},
	volume = {10},
	number = {4},
	year = {2019},
	pages = {471-483},
	month = {October-December},
	doi={10.1109/TAFFC.2017.2736999},
}

@article{Eyben_2016,
  author={F. Eyben and K. Scherer and B. Schuller and J. Sundberg and E. Andr\'{e} and C. Busso and L. Devillers and J. Epps and P. Laukka and S. Narayanan and K. Truong},
  title={The {Geneva} Minimalistic Acoustic Parameter Set ({GeMAPS}) for Voice Research and Affective Computing},
  journal={IEEE Transactions on Affective Computing},
  number={2},
  volume={7},
  pages={190-202},
  year={2016},
  month={April-June},
  doi={10.1109/TAFFC.2015.2457417},
}

@InProceedings{Chang_2024,
	author={X. Chang and B. Yan and K. Choi and J.-W. Jung and Y. Lu and S. Maiti and R. Sharma and J. Shi and J. Tian and S. Watanabe and others},
	title={Exploring Speech Recognition, Translation, and Understanding with Discrete Speech Units: A Comparative Study},
	booktitle={IEEE International Conference on Acoustics, Speech and Signal Processing (ICASSP 2024)},
	volume={},
	year={2024},
	month={},
	address={},
	pages={11481-11485},
	doi={},
	publisher={IEEE},
}

@InProceedings{Nguyen_2023,
	author={T. A. Nguyen and W.-N. Hsu and A. d'Avirro and B. Shi and I. Gat and M. Fazel-Zarani and T. Remez and J. Copet and G. Synnaeve and M. Hassid and others},
	title={Expresso: A Benchmark and Analysis of Discrete Expressive Speech Resynthesis},
	booktitle={INTERSPEECH 2023},
	volume={},
	year={2023},
	month={},
	address={},
	pages={4823-4827},
	doi={},
	publisher={ISCA},
}

@InProceedings{Ren_2024,
	author={W. Ren and Y.-C. Lin and H.-C. Chou and H. Wu and Y.-C. Wu and C.-C. Lee and H.-Y. Lee and H.-M. Wang and Y. Tsao},
	title={Emo-codec: An In-depth Look at Emotion Preservation Capacity of Legacy and Neural Codec Models with Subjective and Objective Evaluations},
	booktitle={Asia Pacific Signal and Information Processing Association Annual Summit and Conference (APSIPA ASC 2024)},
	volume={},
	year={2024},
	month={},
	address={},
	pages={},
	doi={},
	publisher={IEEE},
}

@InProceedings{Polyak_2021,
	author={A. Polyak and Y. Adi and J. Copet and E. Kharitonov and K. Lakhotia and W.-N. Hsu and A. Mohamed and E. Dupoux},
	title={Speech Resynthesis from Discrete Disentangled Self-Supervised Representations},
	booktitle={Proceedings of Interspeech},
	volume={},
	year={2021},
	month={},
	address={},
	pages={},
	doi={},
}

@InProceedings{Wells_2022,
	author={D. Wells and H. Tang and K. Richmond},
	title={Phonetic Analysis of Self-Supervised Representations of English Speech},
	booktitle={Proceedings of Interspeech},
	volume={},
	year={2022},
	month={},
	address={},
	pages={},
	doi={},
}

@article{Mousavi_2024,
	author={P. Mousavi and others},
	title={DASB - Discrete Audio and Speech Benchmark},
	journal={ArXiv e-prints (arXiv:2406.14294)},
	archivePrefix={arXiv},
	eprint={2406.14294},
	primaryClass={cs.SD},
	volume={},
	number={},
	year={2024},
	pages={},
	month={June},
	doi={},
}

@article{Mousavi_2025,
	author={P. Mousavi and others},
	title={Discrete Audio Tokens: More Than a Survey!},
	journal={ArXiv e-prints (arXiv:2506.10274)},
	archivePrefix={arXiv},
	eprint={2506.10274},
	primaryClass={cs.SD},
	volume={},
	number={},
	year={2025},
	pages={},
	month={June},
	doi={},
}

@article{Li_2025,
	author={S. Li and K. Richmond and S. King},
	title={Segmentation-Variant Codebooks for Preservation of Paralinguistic and Prosodic Information},
	journal={arXiv preprint arXiv:2505.15667)},
	archivePrefix={arXiv},
	eprint={2505.15667},
	primaryClass={},
	volume={},
	number={},
	year={2025},
	pages={},
	month={May},
	doi={},
}

@article{Mousavi_2024_2,
	author={P. Mousavi and others},
	title={How Should We Extract Discrete Audio Tokens from Self-Supervised Models?},
	journal={ArXiv e-prints (arXiv:2406.10735)},
	archivePrefix={arXiv},
	eprint={2406.10735},
	primaryClass={cs.SD},
	volume={},
	number={},
	year={2024},
	pages={},
	month={June},
	doi={},
}

@article{Liu_2024,
	author={X. Liu and others},
	title={From Vision to Audio and Beyond: A Unified Model for Audio-Visual Representation and Generation},
	journal={ArXiv e-prints (arXiv:2409.19132)},
	archivePrefix={arXiv},
	eprint={2409.19132},
	primaryClass={cs.MM},
	volume={},
	number={},
	year={2024},
	pages={},
	month={},
	doi={},
}

@article{Yang_2025,
	author={D. Yang and others},
	title={ALMTokenizer: A Low-bitrate and Semantic-rich Audio Codec Tokenizer for Audio Language Modeling},
	journal={ArXiv e-prints (arXiv:2504.10344)},
	archivePrefix={arXiv},
	eprint={2504.10344},
	primaryClass={cs.SD},
	volume={},
	number={},
	year={2025},
	pages={},
	month={April},
	doi={},
}

@article{Yang_2023,
	author={D. Yang and others},
	title={{UniAudio}: An Audio Foundation Model Toward Universal Audio Generation},
	journal={ArXiv e-prints (arXiv:2310.00704)},
	archivePrefix={arXiv},
	eprint={2310.00704},
	primaryClass={cs.SD},
	volume={},
	number={},
	year={2023},
	pages={},
	month={October},
	doi={},
}

@article{Tang_2023,
	author={C. Tang and W. Yu and G. Sun and X. Chen and T. Tan and W. Li and L. Lu and Z. Ma and C. Zhang},
	title={{SALMONN}: Towards Generic Hearing Abilities for Large Language Models},
	journal={ArXiv e-prints (arXiv:2310.13289)},
	archivePrefix={arXiv},
	eprint={2310.13289},
	primaryClass={cs.SD},
	volume={},
	number={},
	year={2023},
	pages={},
	month={October},
	doi={},
}

@InProceedings{Cheng_2024,
	author={Y. Cheng and Y. Li and J. He and R. Feng},
	title={Mixture of Experts for Audio-Visual Learning},
	booktitle={Proceedings of the 38th Conference on Neural Information Processing Systems (NeurIPS 2024)},
	volume={},
	year={2024},
	month={December},
	address={},
	pages={},
	doi={},
	note={},
}

@article{Zhang_2023_ST,
	author={X. Zhang and D. Zhang and S. Li and Y. Zhou and X. Qiu},
	title={SpeechTokenizer: Unified Speech Tokenizer for Speech Large Language Models},
	journal={ArXiv e-prints (arXiv:2308.16692)},
	archivePrefix={arXiv},
	eprint={2308.16692},
	primaryClass={cs.SD},
	volume={},
	number={},
	year={2023},
	pages={},
	month={August},
	doi={},
	note={},
}

@InProceedings{Kumar_2023_DAC,
	author={R. Kumar and P. Seetharaman and A. Luebs and I. Kumar and K. Kumar},
	title={High-Fidelity Audio Compression with Improved RVQGAN},
	booktitle={Advances in Neural Information Processing Systems},
	volume={36},
	year={2023},
	month={},
	address={},
	pages={},
	doi={},
	note={},
}

@article{Defossez_2022,
	author={A. Défossez and J. Copet and G. Synnaeve and Y. Adi},
	title={High Fidelity Neural Audio Compression},
	journal={ArXiv e-prints (arXiv:2210.13438)},
	archivePrefix={arXiv},
	eprint={2210.13438},
	primaryClass={cs.SD},
	volume={},
	number={},
	year={2022},
	pages={},
	month={October},
	doi={},
	note={},
}

\end{document}